\documentclass{article}

\usepackage{PRIMEarxiv}

\usepackage[utf8]{inputenc} 
\usepackage[T1]{fontenc}    
\usepackage{hyperref}       
\usepackage{url}            
\usepackage{booktabs}       
\usepackage{amsfonts}       
\usepackage{nicefrac}       
\usepackage{microtype}      
\usepackage{lipsum}
\usepackage{fancyhdr}       
\usepackage{graphicx}       
\graphicspath{{media/}}     

\title{Extremely Low Thermal Resistance Architectures for AlxGaN1-x Semiconductor Devices
}

\author{
Kidus Guye$^{1}$,
Davide Orlandini$^{2}$,
Seungheon Shin$^{2}$,
Andy Allerman$^{3}$,
Damena Agonafer$^{1}$,
Siddharth Rajan$^{2}$, \AND
Samuel Graham$^{*1}$\\[1ex]
$^{1}$Department of Mechanical Engineering, University of Maryland, College Park\\
$^{2}$Department of Electrical and Computer Engineering, The Ohio State University, Columbus\\
$^{3}$Sandia National Laboratories, Albuquerque\\
\texttt{*samuelg@umd.edu}
}
\begin{document}
\maketitle

\begin{abstract}
Next-generation high-power radio-frequency (RF) devices increasingly demand transistors that operate efficiently with high gain at high frequencies. High-aluminum-content ultra-wide-bandgap (UWBG) AlGaN alloys have shown great potential for enabling such high-frequency RF technologies. However, the widespread adoption of AlGaN-based RF devices is limited by thermal-management challenges arising from the intrinsically low thermal conductivity of AlGaN, which leads to higher device thermal resistance for a given geometry compared to GaN RF devices. As a result, these next-generation devices are highly susceptible to self-heating. This study investigates the thermal behavior of UWBG AlGaN devices, focusing on the effects of AlGaN channel thickness, substrate technology, and high-k material integration on reducing device thermal resistance to enable high-power operation. Experimental results demonstrate a record-low thermal resistance of 3.96 mm$\cdot$K/W when an AlN substrate is employed and the AlGaN channel thickness is reduced to 5 nm. These findings provide valuable insights into mitigating thermal limitations in UWBG devices through device-level engineering and strategic integration of high-k materials.
\end{abstract}


\section{Introduction}
Ultra-wide-bandgap (UWBG) semiconductor devices have emerged as promising candidates to meet the escalating performance demands of next-generation high-power and high-frequency electronics~\cite{klein2025rich}. 
Among emerging materials, aluminum-rich AlGaN alloys have shown strong potential for next-generation radio-frequency (RF) transistors, offering superior high-frequency and high-voltage capabilities~\cite{baca_rf_2019}. This performance advantage arises from their UWBG ($>$5 eV), which enables higher breakdown fields and sustained high electron saturation velocity, resulting in a Johnson figure of merit (JFOM) that surpasses that of GaN-based devices (for example, Al$_{0.85}$Ga$_{0.15}$N exhibits a silicon-normalized JFOM of 81 compared to 30 for GaN)~\cite{shin_high_2025,zhu_heterostructure_2025,baca_-rich_2020,choi_perspective_2021}. However, these electrical advantages are offset by poor thermal performance. The thermal conductivity of AlGaN is up to 21 times lower than that of GaN ($k_{GaN}$ = 180 W/(m$\cdot$K)) for Al compositions between 0.2 and 0.8~\cite{chatterjee2022algan}, which limits heat dissipation and leads to severe self-heating, ultimately affecting device performance and reliability.

Thermal management remains a critical design challenge for high-power semiconductor devices. Prior studies have shown that device-level engineering can reduce thermal impedance through optimized layer architecture, substrate selection, and improved interface quality. For GaN-based wide-bandgap devices (WBG), significant progress has been achieved by reducing channel temperature through buffer-layer and substrate engineering~\cite{kim_thermal_2025,aller_impact_2025,nochetto_impact_2011,schwitter_impact_2014}. Aller \textit{et al}. reported a record-low thermal boundary resistance (TBR) of 3.4 m$^2$K/(GW) across nanocrystalline diamond/AlGaN interfaces, demonstrating the effectiveness of diamond as a heat-spreading layer~\cite{aller_low_nodate}. Tsurumi \textit{et al}. found that replacing conventional SiN passivation with AlN significantly enhanced lateral heat spreading in AlGaN/GaN HFETs~\cite{tsurumi_aln_2010}. Hirama \textit{et al}. achieved a thermal resistance as low as 4.1 K$\cdot$mm/W using single-crystal diamond substrates in AlGaN/GaN High Electron Mobility Transistors (HEMTs), underscoring the benefits of high-thermal-conductivity substrates~\cite{hirama_algangan_2011}. Alam \textit{et al}. demonstrated that incorporating thick AlN heat-spreading layers into Al$_{0.26}$Ga$_{0.74}$N/GaN HEMTs reduced thermal resistance by more than a factor of 2 compared to devices without a heat spreader~\cite{alam_excimer_2021}. Likewise, additional numerical studies showed that double-sided diamond cooling can substantially lower channel temperature in $\beta-$Ga$_2$O$_3$ devices~\cite{kim_transient_2023}. Similarly, a reduction in thermal resistance by a factor of 3 has been reported for AlGaN channel devices using single-crystal AlN substrates compared to AlN/sapphire configurations~\cite{mamun_al064ga036n_2023}. Despite these advances, the thermal resistance of AlGaN devices remains relatively high, and limited research has directly addressed the intrinsic thermal limitations of these low-$k$, high-power UWBG materials. To enable AlGaN alloys for next-generation RF applications, effective thermal management strategies must be developed.\par
In this study, we demonstrate that channel and substrate engineering are important in providing a thermal solution to help mitigate the thermal challenges in AlGaN UWBG devices. Device architectures with varying channel thicknesses are compared to evaluate the effect of channel scaling on thermal resistance and electrical behavior. To further assess the influence of substrate material, identical device architectures are fabricated on both sapphire and AlN substrates. Numerical simulations and experimental measurements are conducted to analyze the impact of channel and substrate design under steady-state and transient operating conditions. The device featuring a thin (5 nm) channel on an AlN substrate exhibited a record-low thermal resistance of 3.96 K$\cdot$mm/W, representing a significant improvement in the heat dissipation capability.

\section{Device Description}
\begin{figure*}[h]
\centering
\includegraphics[scale=0.58]{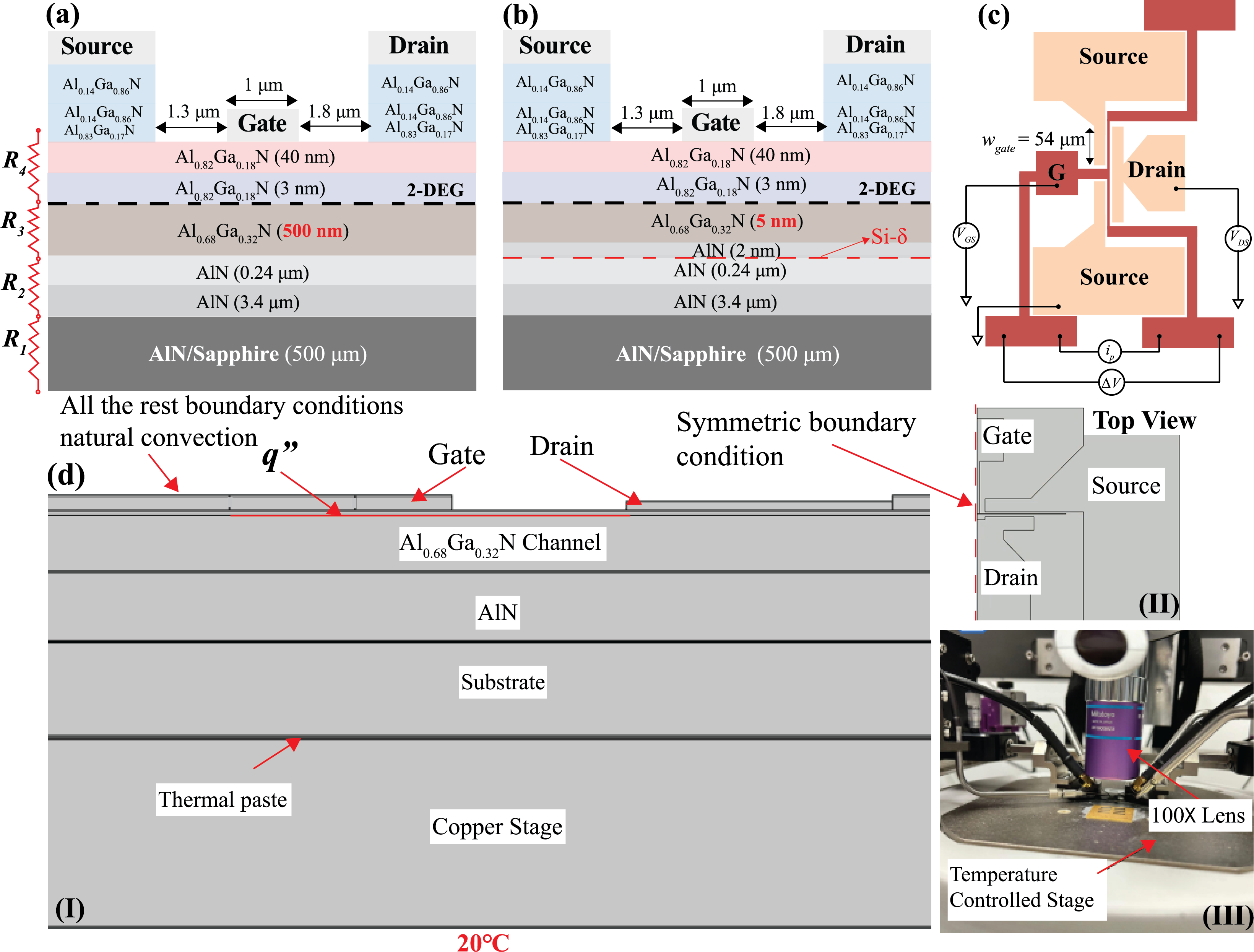}
\caption{\label{Fig1} 
Cross-sectional schematic diagrams of the AlGaN HEMT devices: (a) thick-channel device with a 500~nm channel fabricated on AlN (AlGaN-500-on-AlN) and sapphire (AlGaN-500-on-sapphire) substrates with a simplified thermal resistance networks across the layers where $R_1$ is the resistance across the substrate, $R_2$ is the combined resistance across the buffer layer, $R_3$ is the resistance across the channel, and $R_4$ is the resistance across the barrier layers, and (b) thin-channel device with a 5~nm channel fabricated on AlN (AlGaN-5-on-AlN) and sapphire (AlGaN-5-on-sapphire) substrates. (c) Schematic illustration of the gate resistance thermometry (GRT) design used for thermal characterization. (d) Numerical setup, (I) the schematics the numerical setup with all the boundary conditions, (II) top view of the source, drain and gate in the numerical setup, (III) image of the experimental setup on the SanjSCOPE\texttrademark EZ500 with the 100X objective lens.}
\end{figure*}

\begin{figure}[h]
\centering
\includegraphics[scale=0.3]{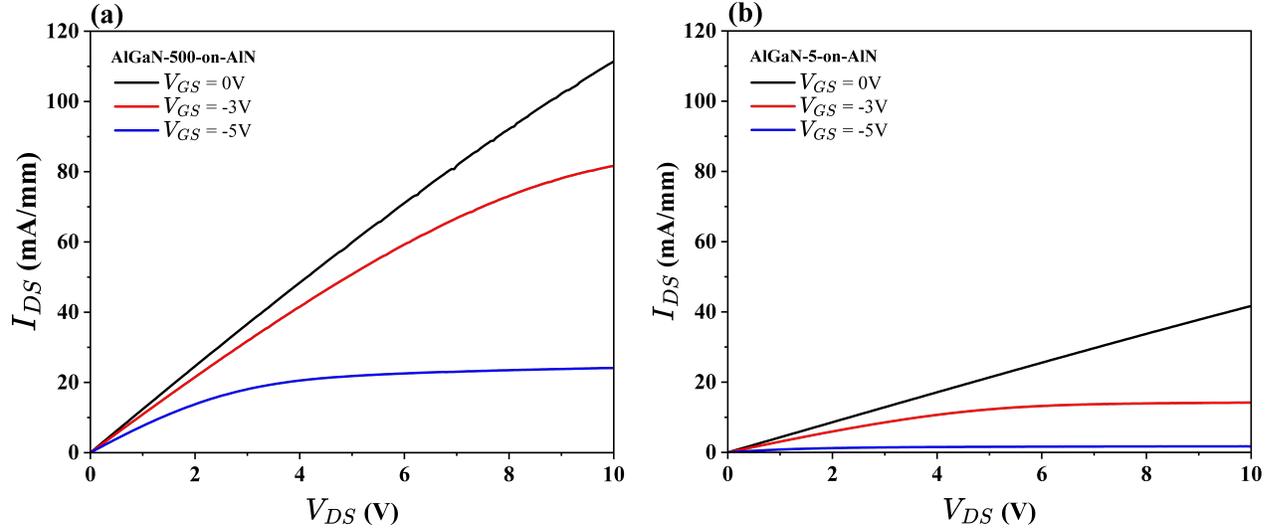}
\caption{\label{FigD1} Output characteristics of the fabricated HFET devices:
(a) a 500 nm AlGaN channel on AlN ( AlGaN-500-on-AlN ) and
(b) a 5 nm AlGaN channel on AlN (AlGaN-5-on-AlN).
For each device, the drain current \(I_D\) was measured by sweeping the drain–to-source voltage \(V_{DS}\) from 0 to 10~V under gate-to–source voltages \(V_{GS}\) of 0 V, -3 V and –5~V.}
\end{figure}

Four two-finger AlGaN heterostructure field effect transistors (HFETs) were fabricated on both sapphire and AlN substrates, featuring AlGaN channel thicknesses of 5~nm (AlGaN-5-on-sapphire and AlGaN-5-on-AlN) and 500~nm (AlGaN-500-on-sapphire and AlGaN-500-on-AlN). The schematic structures of the fabricated AlGaN HFETs with 500~nm channel thickness and 5~nm channel thickness are illustrated, respectively, in Figure~\ref{Fig1}a and~\ref{Fig1}b, with a simplified thermal resistance network across the layers.
Two epitaxial structures were grown on a TNSC-4000HT metal-organic chemical vapor deposition (MOCVD) reactor on both sapphire and AlN substrates. The first epitaxial structure consists, from bottom to top, of a thick AlN buffer layer; a 500~nm Al$_{0.68}$Ga$_{0.32}$N channel layer; a 43~nm Al$_{0.82}$Ga$_{0.18}$N split-doped barrier layer; and a 60~nm reverse-graded contact layer with the Al composition graded from Al$_{0.83}$Ga$_{0.17}$N to Al$_{0.14}$Ga$_{0.86}$N. The exact function and mechanism of the split-doped barrier layer are described in~\cite{Shin_Barrier_2025}.
The second epitaxial structure includes the same epitaxial layers (reverse-graded contact layers, split-doped barrier layer, and buffer layer), but features a 5~nm ultra-thin unintentionally doped (UID) Al$_{0.68}$Ga$_{0.32}$N channel layer.
Four main fabrication steps were performed on both epitaxial wafers to obtain the HFET devices shown in Figure~\ref{Fig1}a and~\ref{Fig1}b: (1) contact layer etching (targeted etching depth of 73~nm), (2) mesa isolation etching (targeted etching depth of 200~nm), (3) source/drain metal deposition (Ti/Al/Ni/Au = 20/120/30/100~nm), and (4) gate metal and gate-resistance thermometry (GRT) structure deposition (Ni/Au = 30/100~nm). \textbf{Figure}~\ref{FigD1}a and \ref{FigD1}b show the measured output characteristics for AlN-substrate devices.\par 

The sheet resistance (R$_{Sh}$), the total sheet charge density (n$_s$), and the 2DEG electron mobility ($\mu_n$) were estimated through Hall-effect measurements performed on Van der Pauw structures fabricated on each sample. The extracted values are summarized in Table~\ref{table1}. 
The mobility and charge in the thin AlGaN-channel devices are lower, possibly due to interfacial impurities or uncompensated interface polarization charge. Recent work on thin-AlGaN channel devices \cite{zhu2025ultrawidebandgapalganheterostructure} suggests that, with appropriate growth conditions, thin buffer layers can still have high mobility ($>$ 200 cm$^2$/Vs).\par

\begin{table}[ht]
\centering
\begin{tabular}{lccc}
\toprule
Sample  &  R$_{Sh}$ ($k\Omega/\square$)  &  n$_s$ (cm$^{-2}$)  &  $\mu_{n}$ (cm$^2$/Vs) \\
\midrule
AlGaN-500-on-sapphire  &   4.16  &  \(1.22 \times 10^{13}\)  &  122 \\
AlGaN-5-on-sapphire    &   8.68  &  \(8.67 \times 10^{12}\)  &   83 \\
AlGaN-500-on-AlN       &   6.19  &  \(8.75 \times 10^{12}\)  &  116 \\
AlGaN-5-on-AlN         &  19.27  &  \(6.42 \times 10^{12}\)  &   50 \\
\bottomrule
\end{tabular}
\caption{Summary of sheet resistance (R$_{Sh}$), total sheet charge density (n$_s$), and 2DEG electron mobility ($\mu_{n}$) extracted from Hall-effect measurements on the four samples.}
\label{table1}
\end{table}

\subsection{Device Thermal Analysis}

To characterize the thermal performance of each device, GRT, an electrical technique that measures temperature based on variations in gate metal resistance, is employed~\cite{pavlidis_gate_2022}. A four-terminal GRT structure is fabricated for each device, as shown in Figure~\ref{Fig1}c. A probe current ($i_p$) is supplied through the GRT metal pads, and the voltage drop across ($\Delta V$) the gate finger is measured using a separate set of probes. The GRT resistance is calibrated over a range of temperatures using a temperature-controlled stage to determine the temperature coefficient of resistance (TCR)~\cite{pavlidis_gate_2022}. A thermocouple was positioned on the die to improve the accuracy of the die temperature measurements during calibration. After calibration, the TCR was extracted from the resistance–temperature relationship and subsequently used to correlate the temperature rise in the gate metal with changes in its electrical resistance. The gate resistance was then measured at different operational power densities, and the corresponding temperature rise in each device was calculated using the extracted TCR. Each device was biased using ground–signal–ground (GSG) probes to apply the drain-to-source voltage ($V_{DS}$) and gate-to-source voltage ($V_{GS}$). A numerical thermal model for each device was developed using COMSOL Multiphysics to compare with the experimental results. A heat-flux boundary condition was applied at the 2-DEG region of the device channel, while a constant-temperature boundary condition of 20~$^\circ$C was imposed at the bottom of the chuck to replicate the experimental stage conditions. All remaining surfaces were assigned natural convection boundary conditions with a heat transfer coefficient ($h$) of 10~W/(m$^2\cdot$K).

\par

\begin{figure}
\centering
\includegraphics[scale=0.25]{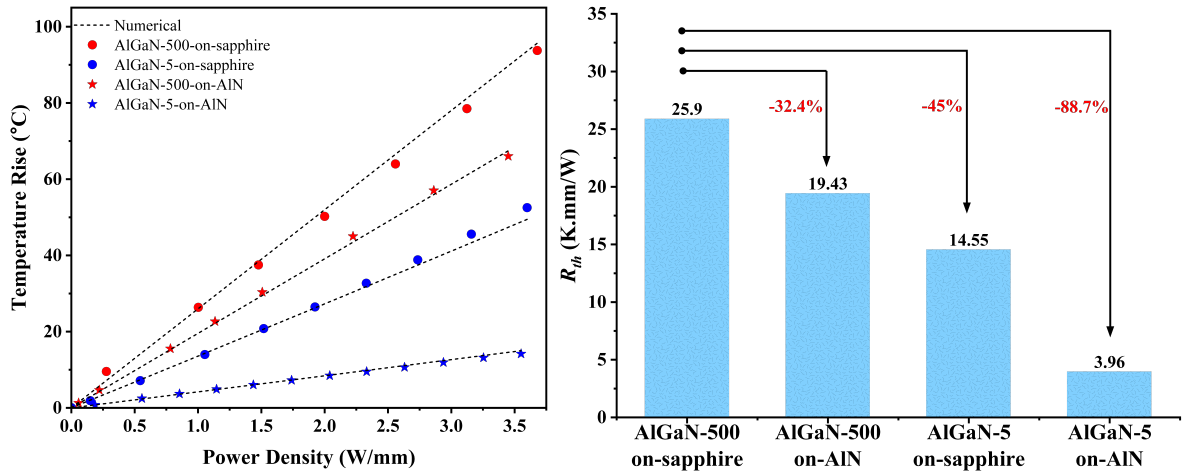}
\caption{\label{Fig2} 
(a) Comparison of the average temperature rise across the gate metal for different power densities for the four different devices with their respective numerical model results. (b) The respective thermal resistance comparison between the four devices.}
\end{figure}
The GRT measurement results for the four devices are shown in Figure~\ref{Fig2}. The data present the average temperature rise across the gate region as a function of power density. To enable a clearer comparison among the different devices, the thermal resistance was calculated, as shown in Figure~\ref{Fig2}b. The 500~nm thick channel device on a sapphire substrate exhibits the poorest thermal performance, with a thermal resistance of 25.9~K$\cdot$mm/W. This degradation is attributed to the low thermal conductivity of sapphire (increased in $R_4$), which limits heat conduction through the substrate to the heat sink, along with increased thermal resistance across the thick channel layer ($R_3$). For the same device architecture, replacing the sapphire substrate with AlN, a high-$k$ material, significantly enhances heat dissipation, reducing the overall thermal resistance by more than 32\%.\par   

Another approach to enhance the thermal performance of AlGaN-based devices is through channel engineering. Owing to the intrinsically low thermal conductivity of AlGaN, reducing the channel thickness effectively decreases the thermal resistance contribution from this layer ($R_3$). As the channel corresponds to the primary heat-generation region, reducing its thermal resistance improves heat spreading and lowers the overall device thermal resistance. Accordingly, the channel thickness was reduced from 500 nm to 5 nm. The results indicate that, on a sapphire substrate, the total thermal resistance decreases by 45\% to 14.55 K$\cdot$mm/W compared to the thick-channel device. Furthermore, combining the thin (5 nm) channel with an AlN substrate yielded a record-low thermal resistance of 3.96 K$\cdot$mm/W, an 88.7\% reduction, representing the lowest value reported for AlGaN devices to date, to the best of the authors’ knowledge.\par
\begin{figure}
\includegraphics[scale=0.25]{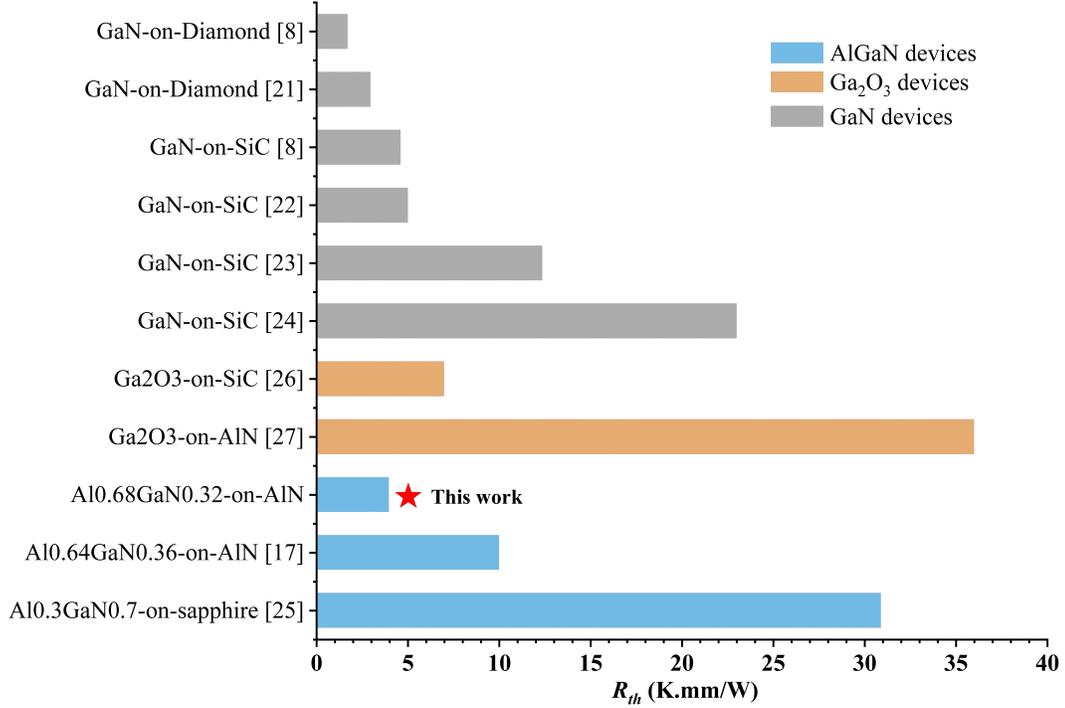}
\centering
\caption{\label{Fig3} 
Comparison of thermal resistance value from this work and with reported work for GaN~\cite{hirama_rf_2012,tadjer_gan--diamond_2019,kim_thermal_2025,pavlidis_thermal_2024,kagawa_high_2024}, AlGaN~\cite{mamun_al064ga036n_2023,chatterjee_interdependence_2020} and Ga$_2$O$_3$~\cite{qu_extremely_2024,lundh_electrothermal_2024} devices.}
\end{figure}

 
Figure~\ref{Fig3} compares the thermal resistance values from this study with previously reported results for GaN and Ga$_2$O$_3$ devices using various thermal management strategies. The results indicate that the thermal resistance achieved in this work is comparable to that of GaN devices on SiC substrates. Furthermore, the recorded thermal resistance exhibits better thermal performance than similar UWBG devices, such as Ga$_2$O$_3$ transistors fabricated on SiC.\par
\subsection{Transient Thermal Response}

To further investigate the transient thermal behavior of the devices, surface temperature mapping is performed using transient thermoreflectance imaging (TTI). The TTI measurements were carried out with a SanjSCOPE\texttrademark EZ500 system using a 100X objective lens. This technique is based on the principle that the reflectivity ($R$) of a material changes for a given change in temperature ($\Delta T$) given by~\ref{eq1}. During calibration, a 530 nm pulsed LED excitation source was employed to record the variation in reflectance as a function of temperature~\cite{pavlidis_transient_2018}. From the calibration, the coefficient of thermoreflectance (C$_{th}$) was determined and subsequently used during measurements to generate spatially resolved temperature maps across the device surface.

\begin{equation}C_{th}=\Delta T \times \left(\frac{\Delta R}{R} \right)^{-1}.\label{eq1}\end{equation}

For the transient study, the devices were biased using an AMCAD pulsed I-V system with a square pulse resulting in a power density of 3.35$\pm$0.15 W/mm, for a duty cycle of 25\% and a total time period of 400~$\mu$s. The results were averaged over 80 s, and the 95\% confidence interval was calculated to estimate the uncertainty. 
\begin{figure}[h]
\includegraphics[scale=0.5]{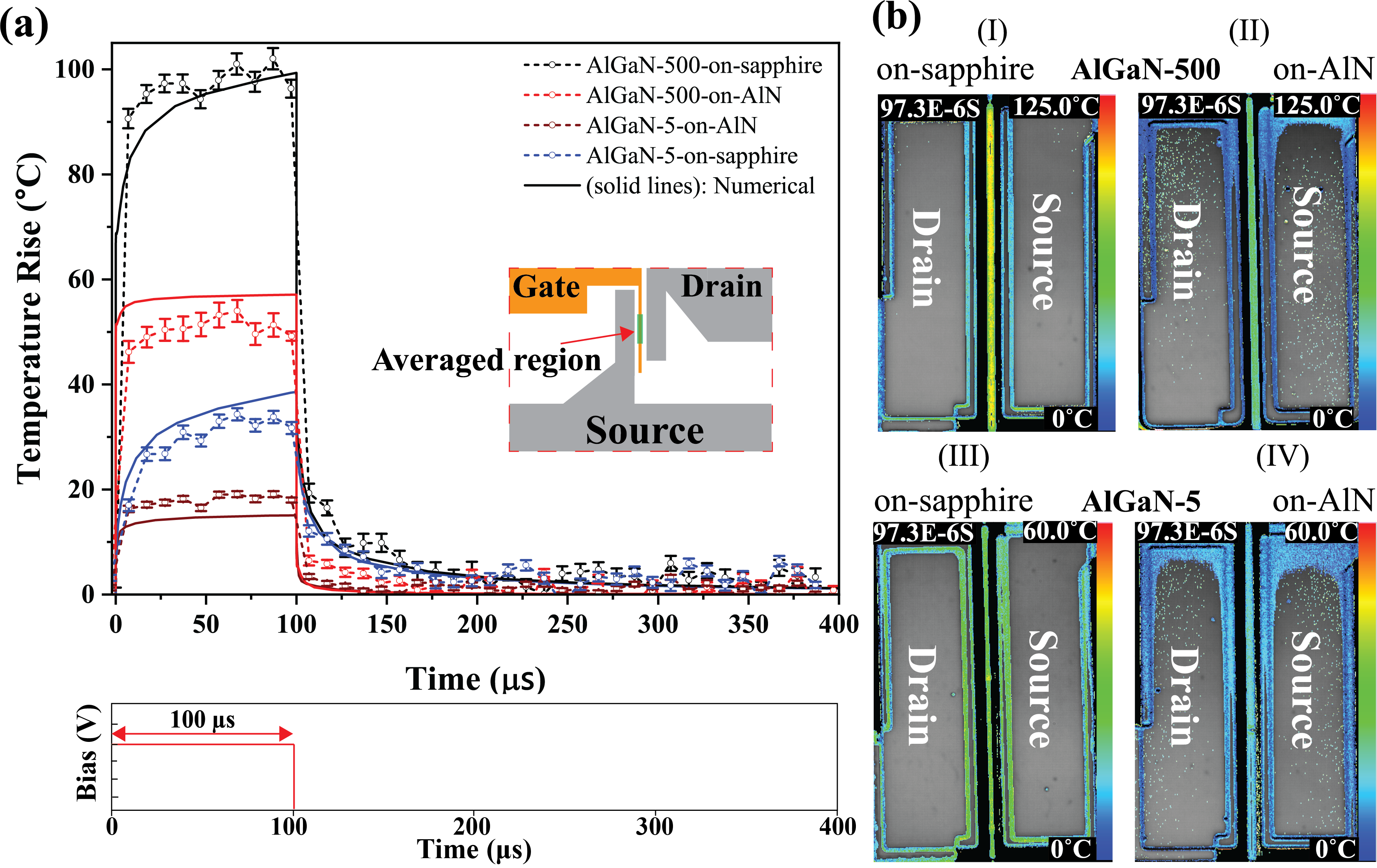}
\centering
\caption{\label{Fig4} 
(a) Transient thermal response showing the temperature rise versus time for the four devices under a 25\% duty cycle with a 400~$\mu$s period at a power density of 3.35$\pm$0.15 W/mm. The solid lines represent the corresponding numerical simulation results for each device. (b) Thermoreflectance imaging illustrating the temperature rise at 97.3 $\mu$s for (I) AlGaN-500-on-sapphire, (II) AlGaN-500-on-AlN, (III) AlGaN-5-on-sapphire, and (IV) AlGaN-5-on-AlN. Note: The color scale for III and IV is reduced to enable clearer comparison.}
\end{figure}

\textbf{Figure}~\ref{Fig4} shows the experimental results, which are compared with numerical simulations performed under the same boundary conditions described earlier. The temperature was averaged across the gate surface near the center region (Figure~\ref{Fig4}a). The data show a clear reduction in temperature rise when transitioning to a thin channel, regardless of the substrate material. The AlGaN-5-on-AlN device exhibits the lowest temperature among all HEMTs, consistent with the steady-state results (Figure~\ref{Fig4}a). This trend is also evident in the thermoreflectance image taken at 97.3 $\mu$s, which highlights the temperature distribution on the gate metal. Figures~\ref{Fig4}b(I)–(II) compare the thick-channel devices on sapphire and AlN substrates, where the sapphire device shows a noticeably higher peak temperature. For the thin-channel devices (Figure~\ref{Fig4}b(III)–(IV)), the temperature contour scale is capped at 60$^\circ$C to enable a more consistent comparison. Both the GRT measurement and the transient study indicate that the thermal challenges associated with using AlGaN as a next-generation RF device material can be mitigated through substrate and channel engineering. In addition, further improvements in interfacial engineering and top-side cooling, such as integrating high-k materials, can provide additional thermal benefits. Although the thin-channel devices exhibited some degradation in electrical performance, they can be further optimized to achieve a balance between thermal management and electrical performance, ensuring overall device optimization~\cite{zhu2025ultrawidebandgapalganheterostructure}.

\section{Conclusion}
This work examines how substrate and channel engineering can address the thermal challenges limiting the adoption of AlGaN-based RF devices. Integrating a high-k material such as AlN as the substrate, together with a thin 5-nm AlGaN channel, reduces the overall device thermal resistance by more than 88.7\% compared to a 500-nm-thick AlGaN channel on sapphire. The record-low thermal resistance achieved in AlGaN devices is also comparable to that of GaN devices grown on SiC substrates. These findings demonstrate that through optimized substrate, interfacial, and channel engineering, AlGaN RF devices can be designed to deliver improved thermal performance suitable for high-power and high-frequency operation.

\medskip
\section{Acknowledgments} 
This work was funded by ARO DEVCOM under the Grant No. W911NF2220163 (UWBG RF Center, program manager Dr. Tom Oder).
\medskip

\medskip
\textbf{Conflict of Interest} \par 
The authors declare no conflict of interest.

\medskip

\bibliographystyle{unsrt}  
\bibliography{references}

\end{document}